\documentclass[twocolumn]{aastex631}

\def\beq{\begin{equation}}
\def\eeq{\end{equation}}

\def\a{{\alpha}}
\def\b{{\beta}}

\def\s{{\it Swift}}

\usepackage{graphicx}
\usepackage{enumitem}
\usepackage{pgfplots}
\usepackage{booktabs}
\pgfplotsset{compat=1.18}

\usepackage{siunitx} 

\usepackage{tikz}
\usepackage{pgfplots}
\pgfplotsset{compat=1.18} 
\usepackage{float}
\usepackage{subfigure} 
\begin{document}

\title{Cosmological Evolution of Gamma Ray Bursts}

\author{Sujay Champati}
\affiliation{California Institute of Technology, Pasadena, CA 91125, USA;
\underline{sujay@caltech.edu}}

\author{Vah\'e Petrosian}
\affiliation{Department of Physics and KIPAC, Stanford University Stanford, CA 94305, USA; \underline{vahep@stanford.edu}}

\affiliation{Department of Applied Physics, Stanford University, Stanford, CA 94305, USA}

\author{Maria G. Dainotti}
\affiliation{National Astronomical Observatory of Japan, Mitaka, Tokyo 181-8588, Japan; \underline{mariagiovannadainotti@yahoo.it}}

\affiliation{The Graduate University for Advanced Studies, SOKENDAI, Kanagawa 240-0193, Japan}

\affiliation{Space Science Institute, Boulder 80301, CO, USA}

\begin{abstract}

Gamma-ray bursts (GRBs) are classified as long (LGRBs) and short (SGRBs), with collapsars and compact-object mergers (NS–NS or NS–Black Holes) as progenitors, respectively. LGRBs are expected to follow the cosmic star formation rate (SFR), while SGRBs follow a delayed version of the SFR. However, this division has come under question, most prominently by observational evidence of an excess of LGRBs at low redshifts by several investigations, summarized in \cite{Petrosian_2024}. Two recent observations of low-redshift LGRBs show associations with kilonovae. Both of these indicate compact mergers as a potential source of LGRBs as well. Most results showing this separation are based on analyses of small (less than 200) samples of LGRBs with measured redshifts. The aim of this paper is to use a larger sample of LGRBs. The number of LGRBs with measured redshifts has increased by more than a factor of 2 over the last decade. To this data set we add a sample of LGRBs whose redshifts are estimated using a machine learning (ML) method (\cite{Narendra_2025}). To account for the observational selection bias due to redshift measurements, we use the non-parametric, non-binning Efron-Petrosian method to establish the degree of correlation between luminosity and redshift, \textit{the luminosity evolution}, and then use the  Lynden-Bell $C^-$  method to obtain the luminosity function. We find a low redshift excess for the larger sample with measured redshifts. Adding the sources with ML-estimated redshifts, which shows overabundance of  in the mid-range redshifts, the excess is reduced.
\end{abstract}


\section{Introduction}
\label{sec:intro}

Gamma-ray bursts (GRBs) are the most luminous electromagnetic transients in the universe, with isotropic-equivalent luminosities reaching up to $10^{52} \,\mathrm{erg/s}$ \citep{Yu_2022}. Discovered in 1967 and first reported by \citet{1973ApJ...182L..85K}, they can be detected across cosmological distances, with some observed beyond redshift $z \sim 9$ \citep{Cucchiara_2011, Salvaterra_2009}. Their extraordinary brightness makes them not only valuable probes of relativistic astrophysics but also powerful beacons for studying the early Universe, tracing star formation, metallicity evolution, and reionization \citep{Lamb_2000, Totani_1997}.  

GRBs are broadly divided into two classes: long-duration GRBs (LGRBs), associated with the collapse of massive stars \citep{1993ApJ...405..273W, hjorth2011gammarayburstsupernova}, and short GRBs (SGRBs), attributed to compact binary mergers \citep{2014ARA&A..52...43B}. LGRBs are naturally expected to track the cosmic star formation rate (SFR). SGRBs, by contrast, may trace older stellar populations on longer delay timescales, but they contribute less to the overall GRB population at high redshifts.%
\footnote{This division, first established in \cite{1993ApJ...413L.101K}, is based on the $T_{90}$ duration, the length of the period containing 90\% of the counts, with separation at $T_{90}=2$s.}
However, this division has been challenged by several recent results. Multiple independent studies report an excess of LGRBs at low redshifts compared to expectations from the SFR \citep{Petrosian_2024}. Moreover, $\mathrm{GRB}211211\mathrm{A}$ \citep{Rastinejad2022Natur.612..223R} and $\mathrm{GRB}230307\mathrm{A}$ \citep{Levan2024,Yang2024} have recently shown association with kilonovae, suggesting that a subset of LGRBs may share progenitors with SGRBs. Another study, \cite{Dimple_2023}, found evidence of two distinct populations of GRBs associated with kilonovae. These findings motivate a reassessment of GRB formation channels and their evolutionary history, which in turn requires large, well-characterized samples spanning a broad redshift range. 

The \textit{Neil Gehrels Swift Observatory} (\s, hereafter) launched in 2004, has been pivotal in building such samples. Its Burst Alert Telescope (BAT) enables rapid detections, while follow-ups by the X-Ray Telescope (XRT) and Ultraviolet/Optical Telescope (UVOT) provide precise localizations and redshift measurements \citep{2004ApJ...611.1005G}.  Together with The Large Area Telescope (LAT) and Gamma-ray Burst Monitor (GBM) of \textit{Fermi} \citep{2009ApJ...702..791M}, \s~ has established the most comprehensive GRB catalog to date, forming the backbone of population and Giuseppe evolution studies.


Progress has been limited because redshift is secured for a small fraction $(<50\%)$ of detected GRBs, leading to $\lesssim 200$ sources in earlier studies. This introduces some uncertainty on the completeness of the samples. However, the situation is  improving. First, the number of spectroscopically measured redshifts has more than doubled in the past decade. Second, advances in ML  provide reliable redshift estimates for GRBs  without spectroscopic follow-up \citep{Narendra_2025, manchanda2025gammarayburstlightcurve, dainotti2025grbredshiftclassifierfollowup}, reducing the above uncertainty. 

A central challenge is to disentangle the effects of intrinsic luminosity evolution from changes in the GRB formation rate. The luminosity function (LF), which characterizes the distribution of intrinsic GRB luminosities, plays a key role in this effort. Using non-parametric techniques such as the Efron–Petrosian (EP) method \citep{Petrosian_2015}, one can measure luminosity evolution directly from the data and construct bias-corrected luminosity functions via the Lynden-Bell $C^-$ method.  

In this work, we combine the most comprehensive set of \textit{Swift} LGRBs, augmented by ML-based redshift estimates, to investigate luminosity evolution and the formation rate in a larger, less biased sample. We test scenarios of pure luminosity evolution, pure density evolution, and hybrid models, and compare the inferred GRB formation history with the cosmic SFR. This allows us to place updated constraints on LGRB progenitors, particularly the nature of the low-redshift LGRB population, and to demonstrate the power of ML-augmented datasets for probing the role of GRBs in cosmic history.

In the next section we describe the data we use, and in \S 3 we present a brief review of the EP and the $C^-$ methods. The results are presented in \S 4 and in \S 5 we  give a summary and describe our conclusions.

\section{Data} 
\label{sec:data}

The spectroscopic redshifts we use in this analysis come from the \textit{Swift} observatory, which has cataloged 453 GRBs with  measured redshift, with many events having multiple observations by different instruments. Of these 453 events, we discard 9 redshifts due to the presence of only an upper or lower bound on $z$. The population of inferred redshifts by ML from \cite{dainotti2025grbredshiftclassifierfollowup} consist of 276 LGRBs. Deleting sources with $z>14.4$,  the redshift of the furthest galaxy from Earth \citep{naidu2025cosmicmiracleremarkablyluminous} known today, we are left with a total of $251$ LGRBs. 

\begin{figure}[htbp]
  \centering
  \includegraphics[width=0.99\linewidth]{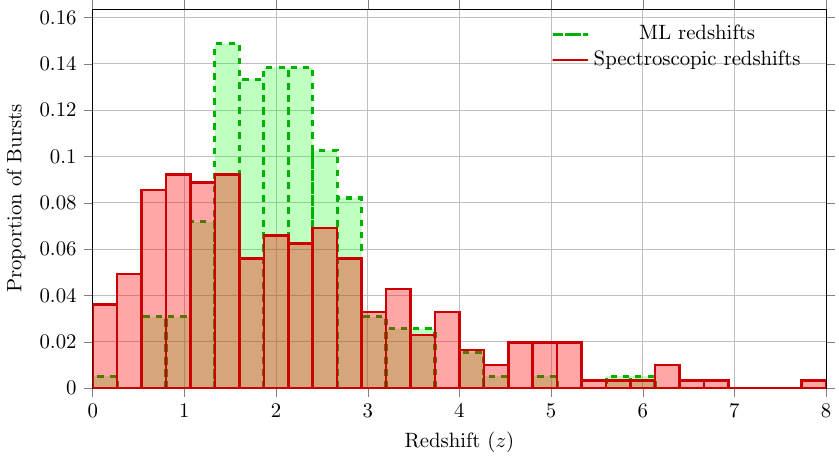}
  \caption{Comparison of fractional binned redshift distributions of ML (green) and non-ML (red) samples.} 
  \label{fig:redshift-hist}
\end{figure}

The union of these two sets gives us a total of $695$ bursts. In Figure \ref{fig:redshift-hist}, we compare the redshift distributions of the two samples. There is a clear difference between these distributions, with the majority of the ML sample in the range $~1.5<z<3$.
As explained in \citep{Narendra_2025}, this discrepancy is due to the fact that the distribution of the variables for the training set, such as $\log NH$, the best predictor, and peak flux, the second best, for the LGRBs with and without redshifts are different. 
The Kolmogorov Smirnov (KS) test  for the first yields a p-value of  $p=0$ and the second $p=0.19$. 
It is then expected that the predicted redshift distributions will be different from the spectroscopic ones. For further information on this discussion, see Fig. 14 in \citep{Narendra_2025}.

For determination of the evolution of the LF, in addition to redshift, we need measured fluxes and gamma-ray spectra. We use the peak energy flux and rest frame luminosity in the 15-150 KeV band. The third \textit{Swift}/BAT catalog has fluxes for 584 of the above sample. For this sample we calculate the peak luminosity for an assumed cosmological model, as%
\footnote{We use the flat $\Lambda$CDM model with matter density parameter $\Omega_m=0.3$ and Hubble constant $H_0=70$ km/s/Mpc.}
\vspace{-0.1cm}
\begin{equation}\label{LofZ}
        L (Z)= 4\pi  d_L^2(Z) \times f(Z)/{K}(Z),
\end{equation}
where $d_L(z)$ is the luminosity distance and $K(z)$ is the \textit{K}-correction factor that accounts for the cosmological redshifting of the observed spectrum. For this we need the gamma-ray spectra.
The \textit{Swift}/BAT Catalog has spectra for 426 bursts in the sample. The simple power law (PL) model  
\vspace{0.1cm} 
\begin{equation}
f(E) = N \left( \frac{E}{E_{\rm norm}} \right)^{\alpha},
\end{equation} 
suffices for a majority of these bursts, while for a few a power-law with high energy exponential cut off (CPL) model  
\begin{equation}
f(E) = N \left( \frac{E}{E_{\rm norm}} \right)^{\alpha} \exp\left(-\frac{E(2+\alpha)}{E_{\rm peak}}\right),
\end{equation}
provides better fit based on  the Sakamoto criterion, $\Delta\chi^2 \equiv \chi^2_{\mathrm{PL}}-\chi^2_{\mathrm{CPL}} > 6$. 
Here $N$ is the normalization (
keV cm$^{-2}$ s$^{-1}$ keV$^{-1}$), $\alpha$ is the spectral index, $E_{\rm peak}$ is the peak energy of the $Ef(E)$ spectrum, and $E_{\rm norm}$ is a reference energy (fixed at 50 keV in our fits).
$74$ bursts in the catalog do not have an associated best fit model, on account of different fitting functions being inconsistent with each other. Further details about the fits to each model are given in \cite{Lien_2016}.

\begin{figure*}
    \centering
    \includegraphics[width=0.7\linewidth]{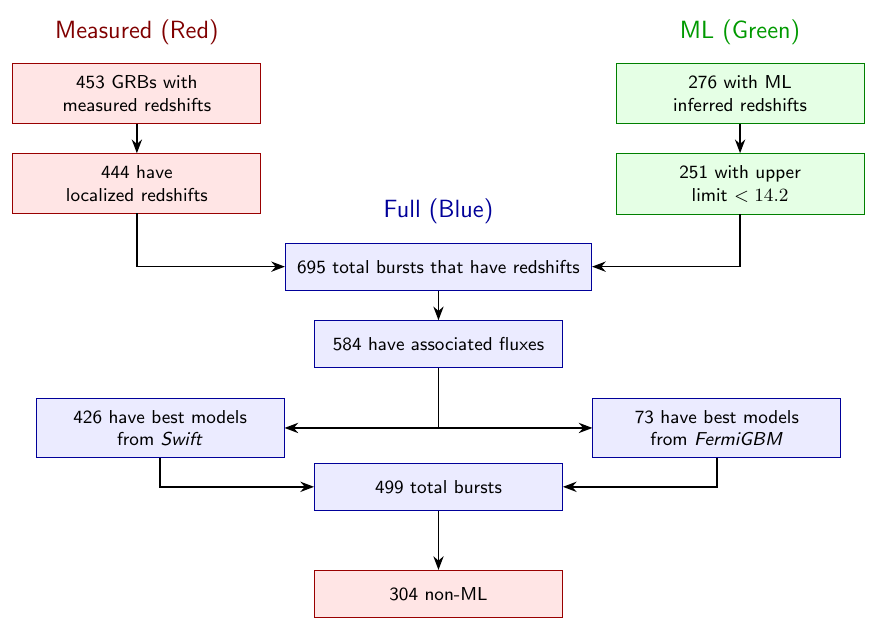}
    \caption{Flowchart of Data Processing Pipeline}
    \label{fig:flowchart}
\end{figure*}

To increase the size of our catalog, we cross-reference the sample with the \textit{Fermi} GBM  Catalog (\cite{Paciesas_2012}). We find a total of $73$ bursts with power law or cutoff power law fits that were undetermined in the \textit{Swift} catalog, bringing our total to $499$ bursts, and $304$ spectroscopic $z$ bursts. The details of this pipeline are shown in Figure \ref{fig:flowchart}.
The $K$-correction translates the observed flux in the instrument’s fixed observer-frame band $[E_1, E_2]$ (15–150 keV for \textit{Swift}/BAT) to the corresponding flux in the burst’s rest frame. For a GRB at redshift $z$, the $K$-correction is given by

\vspace{-0.1cm}
\begin{equation}
K(z) = \frac{\int_{E_1}^{E_2} E \, f(E) \, dE}{\int_{E_1(1+z)}^{E_2(1+z)} E \, f(E) \, dE}, 
\end{equation}
where $f(E)$ is the energy flux spectrum (in units of 
erg cm$^{-2}$ s$^{-1}$ keV$^{-1}$) in the observer frame. The numerator integrates over the desired rest-frame energy band, while the denominator integrates over the redshifted observer-frame band. 

\vspace{0.2cm}


The truncation limit is then defined as:

\beq
L_{\rm min}(Z) = 4\pi d^2_L(Z)f_{\rm lim}/{\bar K}(Z)
\label{Lmin}
\eeq

To compute this limit, we pick $f_\mathrm{lim}$ as the limiting flux of our catalog. As done in \cite{Petrosian_2015}, we choose this limit to be more conservative than the nominal survey threshold, thereby reducing the impact of incompleteness near the detection boundary. We note that the true detectability of a GRB by \textit{Swift}/BAT is not determined by a single flux threshold alone, but can also depend on spectral hardness, burst duration, background conditions, and triggering details. However, the EP method requires an effective truncation boundary in the luminosity-redshift plane in order to define the associated sets. Our adopted conservative flux limit is therefore intended to define a region of the catalog where the sample is approximately complete, rather than to model the full multidimensional detector selection function.

Additionally, we have to compute the function $\bar K(Z)$, which is the average $K-$correction for a given redshift. Since we are using two different models for our $K-$corrections, and due to the fact that values of $K$ can have large variance even for a fixed $Z$, we cannot compute $\bar K(Z)$ directly. 

\begin{figure}[!htbp]
  \centering
  \includegraphics[width=0.99\linewidth]{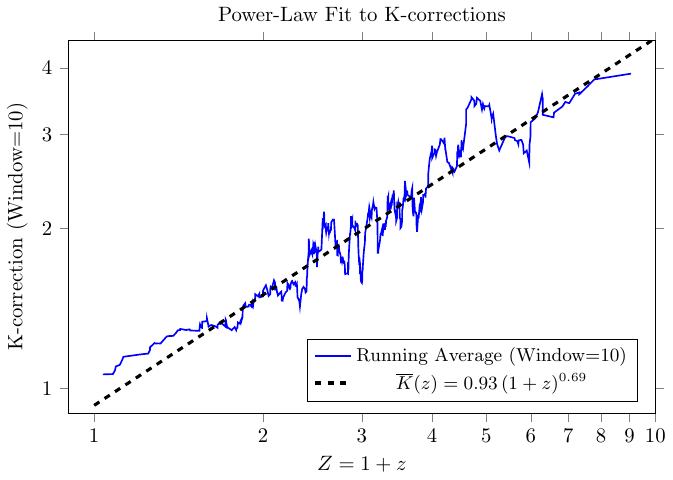} 
  \caption{Power-law fit overlaid on running-average $K$-corrections.}
  \label{fig:kcorrs_bestfit}
\end{figure}

Instead, we construct $\bar{K}(z)$ by applying a running average over all $K$ values sorted by redshift, starting from the lowest $z$ in the sample. For each redshift step, we average over a moving window of 10 neighboring bursts, which smooths fluctuations arising from spectral diversity and statistical noise. This procedure yields a more well-behaved, continuous function $\bar{K}(z)$ that captures the redshift-dependent trend of the $K$–correction while minimizing the impact of outliers or sparsely sampled regions. We then fit a simple power law to this function and use this as our average. The resulting $\bar{K}(z)$ is then used in our luminosity calculations to ensure a consistent and unbiased treatment of the flux limit across the entire LGRB sample. The windowed $K-$corrections and power law fit are shown in Figure \ref{fig:kcorrs_bestfit}.

To assess whether the EP results depend sensitively on the fitted form of $\bar K(Z)$, we performed an additional robustness test. The individual burst luminosities are always computed using their corresponding per-burst $K$-corrections, while $\bar K(Z)$ enters only through the flux-limit boundary in Equation~\ref{Lmin}. In principle, one could construct a boundary using the individual $K$-corrections directly. However, because the individual $K$ values show substantial burst-to-burst scatter, this would introduce sharp local fluctuations into the truncation curve. Such fluctuations can affect the associated sets used in the EP rank statistic in a way that reflects spectral variation rather than the smooth survey selection boundary.

We therefore use the fitted average $\bar K(Z)$ as the baseline truncation boundary, and then test the sensitivity of this choice by perturbing $\bar K(Z)$ within the observed scatter of the smoothed $K$-corrections about the best-fit power law. For each Monte Carlo realization of the perturbed boundary, we recomputed the EP statistic $\tau(k)$ and the corresponding best-fit luminosity-evolution parameter. These perturbations provide a direct test of whether reasonable variations in the average $K$-correction would change the inferred luminosity evolution.

\begin{figure*}[t]Giuseppe 
  \centering
  \includegraphics[width=0.95\textwidth]{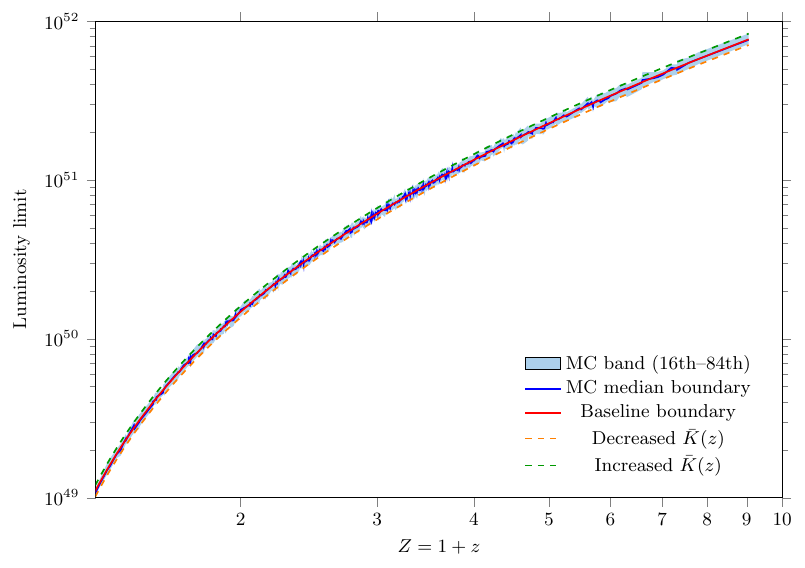}
  \caption{
  Sensitivity of the EP luminosity-limit boundary to perturbations in the fitted average $K$-correction. 
  The baseline boundary is constructed using the power-law fit to the running-average $\bar K(Z)$, while the shaded band shows the $16$th--$84$th percentile range from Monte Carlo perturbations of $\bar K(Z)$ within the measured log-space scatter. 
  The median perturbed boundary and representative systematic perturbations remain close to the baseline curve, indicating that the associated sets used in the EP statistic are not strongly affected by reasonable variations in the adopted smooth representation of $\bar K(Z)$.
  }
  \label{fig:kcorr-perturbations} 
\end{figure*}

As shown in Figure~\ref{fig:kcorr-perturbations}, the perturbed boundaries closely track the baseline boundary, and the inferred luminosity-evolution parameter remains stable. The Monte Carlo perturbations give $k = 2.867 \pm 0.072$, consistent with the baseline value $k = 2.900$ obtained from the fitted average $\bar K(Z)$. Thus, the inferred luminosity evolution is not strongly sensitive to the adopted smooth representation of the average $K$-correction.

Furthermore, from Figure \ref{fig:redshift-hist}, it is evident that the distributions of the two samples vary considerably in their spread. In order to determine whether the ML redshifts could feasibly be drawn as a sample from the spectroscopic distribution, we apply the Anderson-Darling test, which yields $p=0.001$ confirming that these samples are indeed distributed differently. We thus conduct our analysis on both the combined sample and the spectroscopic sample, and comment on the differences in section \ref{sec:concls}.

Finally, as a further check on the ML-estimated redshifts, we considered the small subset of GRBs for which both spectroscopic and ML-predicted redshifts are available. This overlap sample contains only about 20 GRBs, and is therefore too small to draw statistically significant conclusions about the luminosity evolution or formation-rate history from this subset alone. We therefore do not treat this overlap sample as an independent catalog for the main analysis. However, we note that a previous tests using {\it the statistical-learning redshift method} by \cite{2024ApJ...967L..30D} obtained ML-inferred redshift for a complete sample of 171 GRBs with known  spectroscopic redshifts and observed plateaus. Comparison of the two redshift samples shows a Pearson correlation coefficient of $0.93$ between the inferred and observed redshifts, and  similar luminosity evolution index $k$, luminosity function and density rate evolution for both redshift samples.

We return to the implications of this limitation when discussing the ML-augmented density rate evolution below.

\section{Methods and Approach} \label{sec:methods}

Our goal is to reconstruct the bivariate distribution $\Psi(L, Z)$, while correcting for observational truncation (i.e., Malmquist or Eddington bias). Traditional approaches often use forward fitting modeling, assuming specific parametric forms for $\Psi(L, Z)$ and fitting to observed data \cite{Petrosian1992}. In contrast, nonparametric techniques such as $V/V_{\rm max}$ \cite{1968ApJ...151..393S} and the $C^{-}$ method \cite{1971MNRAS.155...95L}%
\footnote{The $C^{-}$ method was independently rediscovered by \cite{9ea8269a-ec5c-3ba1-9f46-7c7bedadb11c} and \cite{a3bacb85-0f9c-343f-9e39-06add35203dd}, but we use the astrophysical naming convention.} 
infer distributions directly from the data. 

However, as emphasized in \cite{Petrosian1992}, these nonparametric methods assume statistical independence between $L$ and $Z$, i.e., $\Psi(L,Z) = \psi(L)\rho(Z)$, and thus cannot capture luminosity evolution. To address this, Efron and Petrosian (EP) developed a method \cite{1992ApJ...399..345E} that tests for correlations between variables under one-sided truncation.%
\footnote{EP later extended this approach to handle two-sided truncation in \cite{Efron01091999}.}
\begin{figure*}[]
    \centering
    \includegraphics[width=0.48\linewidth]{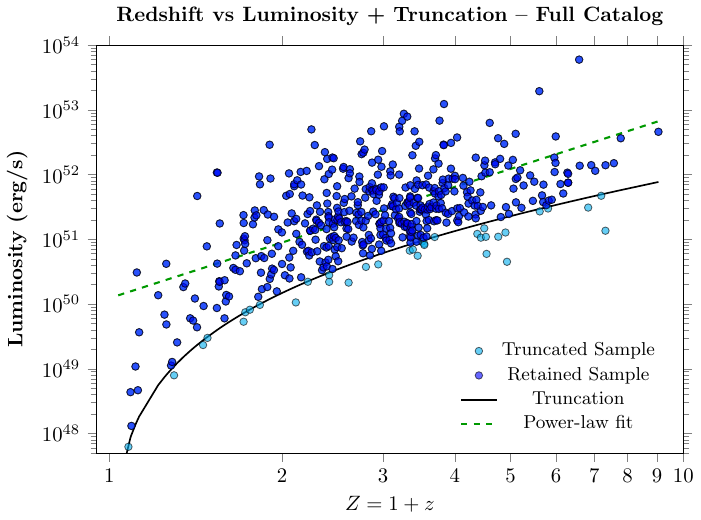}
    \includegraphics[width=0.48\linewidth]{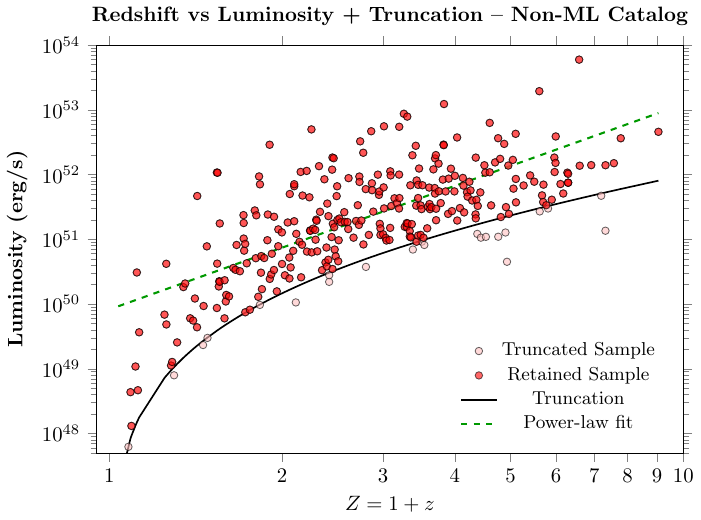}
    \caption{({\it Left}): Redshifts and Luminosities for combined GRB catalog. The black truncation line is obtained from $f_{\mathrm{lim}}$ and the average $K$. The power law fit for the bursts is shown in green. ({\it Right}): Redshifts and Luminosities for the Non-ML Catalog}
     \label{fig:scatters}
\end{figure*}
We adopt the EP procedure to provide a quantitative description of  luminosity evolution and introduce a transformed variable $L_0 = L(Z)/g(Z)$, with parameters of the evolution function $g(Z)$ chosen such that $L_0$ is uncorrelated with $Z$ once $L_0$ and $Z$ are rendered independent.

This methodology has been widely applied to the analysis of AGNs and GRBs (see citations above). It proceeds by constructing an associated set for each data point $(Z_i, L_i)$ using the survey's detection limits: $L_{\min}(Z_i)$ and $Z_{\max}(L_i)$. Each data point is ranked within its associated set based on either $L_i$ or $Z_i$. The associated set for a point includes all sources $(Z_j, L_j)$ satisfying either $Z_j \leq Z_i$ and $L_j \geq L_{\min}(Z_i)$, or $L_j \geq L_i$ and $Z_j \leq Z_{\max}(L_i)$.

The degree of correlation is quantified using Kendall’s Tau statistic:
\begin{equation}
    \tau = \frac{\sum_i(R_i - E_i)}{\sqrt{\sum_i V_i}},
\end{equation}
where $R_i$ is the rank of $(L_i, Z_i)$ in its associated set, and $E_i = \frac{N_i + 1}{2}$, $V_i = \frac{N_i^2 - 1}{12}$ are the expected mean and variance for a set of size $N_i$. For independent variables, we expect $\tau = 0$; the significance of any deviation quantifies the degree of correlation. 

To model the evolution, we use a flexible broken power-law form for $g(Z)$:
\begin{equation}
    g(Z) = Z^k \frac{1 + Z_{\rm cr}^k}{Z^k + Z_{\rm cr}^k},
    \label{eq:Lumfit}
\end{equation}
which approaches a constant at high redshifts, reflecting the flattening of the cosmic expansion rate. The function is normalized such that $g(1) = 1$, making the de-evolved $L_0$ interpretable as local luminosity. This form prevents the inferred evolution rate from exceeding $H(Z)$, which flattens near $Z \sim 3$–4 in $\Lambda$CDM cosmology. Based on previous studies (e.g., \cite{Petrosian_2015}), we adopt $Z_{\rm cr} \approx 3.5$. The optimal value of $k$ is the one that yields $\tau = 0$, and the $1\sigma$ uncertainty is defined by the range of $k$ where $\tau = \pm 1$.

With $k$ determined, we apply the $C^{-}$ method to derive the cumulative local luminosity function:
\begin{equation}
    \phi(L_0) = \int_{L_0}^\infty \psi(L')\, dL',
\end{equation}
and the cumulative source co-moving number rate:
\begin{equation}
    \dot\sigma(Z) = \int_1^Z \frac{\dot\rho(Z')}{Z'} \frac{dV(Z')}{dZ'}\, dZ',
\end{equation}
where $\dot\rho(Z)$ is the co-moving formation rate density, and $V(Z)$ is the co-moving volume up to redshift $Z$.

To evaluate $\phi(L_0)$, we sort observed luminosities in descending order ($L_1$ being the highest), and use:
\begin{equation}
    \phi(L_j) = \phi(L_1) \prod_{i=2}^j \left(1 + \frac{1}{N_i} \right),
\end{equation}
where $N_i$ is the size of the associated set for source $i$, defined by $L_j \geq L_i$ and $Z_j \leq Z_{\max}(L_i)$. $\phi(L_1)$ denotes the cumulative function above the brightest observed luminosity.

A similar approach yields the cumulative redshift distribution. Starting from the lowest redshift $Z_1$, we compute:
\begin{equation}
    \dot\sigma(Z_j) = \dot\sigma(Z_1) \prod_{i=2}^j \left(1 + \frac{1}{M_i} \right),
\end{equation}
where $M_i$ is the size of the associated set for source $i$, defined by $Z_j \leq Z_i$ and $L_j \geq L_{\min}(Z_i)$. $\dot\sigma(Z_1)$ denotes the cumulative rate between $Z = 0$ and the lowest observed redshift.

\begin{figure}[h!]
    \centering
    \includegraphics[width=0.99\linewidth]{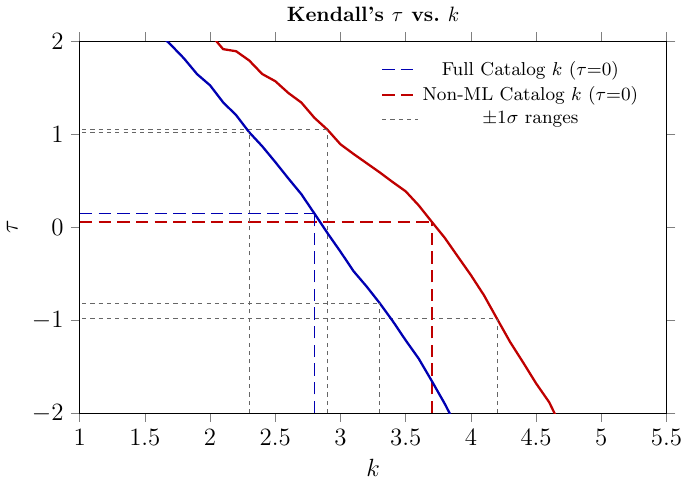}
    \caption{
    Kendall’s $\tau$ statistic as a function of $k$ for the full GRB catalog (blue) and the non-ML subset (red). 
    Solid curves show the measured $\tau(k)$. 
    Dashed lines indicate the central (fit) guides for each dataset, while the lighter dashed bands denote the corresponding $\pm 1\sigma$ ranges. 
    }
    \label{fig:kvstau}
\end{figure}

\begin{figure*}[]
    \centering
    \includegraphics[width=0.48\linewidth]{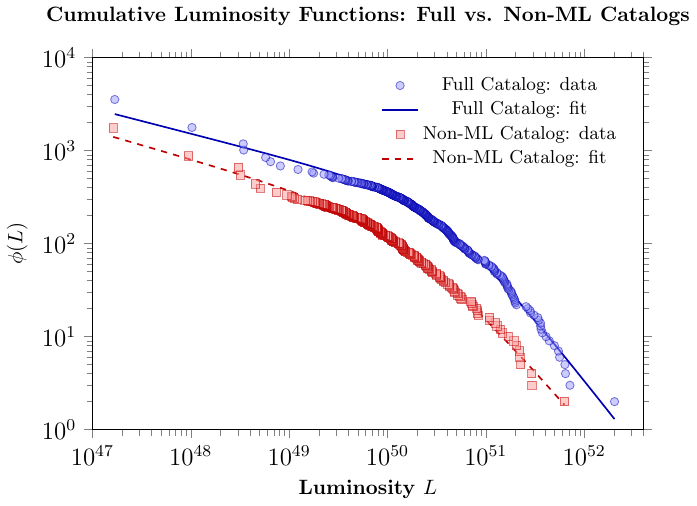}
    \includegraphics[width=0.48\linewidth]{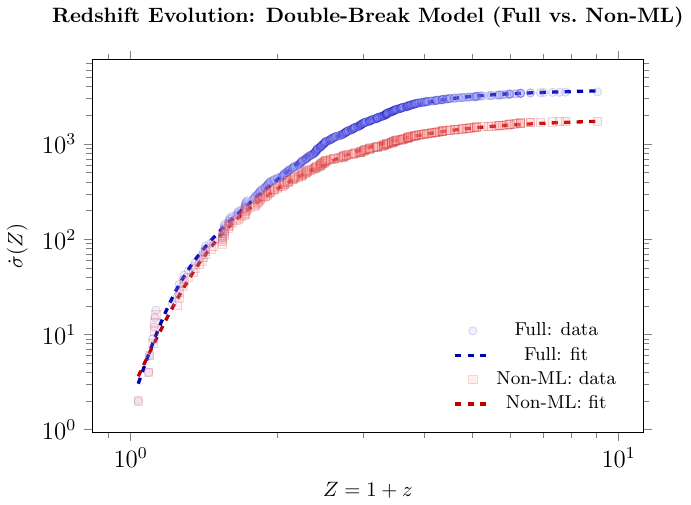}
    \caption{({\it Left}): Formation rates derived from the $C^-$ method for both the full (blue) and non-ML (red) GRB catalogs. The solid curves indicate the best-fit double-break power law models. Cumulative luminosity functions for the GRB catalogs. 
    Blue points with a solid fit line represent the full catalog, while red squares with a dashed fit line represent the non-ML subset. 
    Both are fitted with a smooth double power law model. ({\it Right}): Formation rates derived from the $C^-$ method for both the full (blue) and non-ML (red) GRB catalogs. The solid curves indicate the best-fit double-break power law models.}
     \label{fig:lumsforms}
\end{figure*}

\newpage

\section{Results}

\subsection{Luminosity Evolution}

To test for independence, we apply the Efron-Petrosian Methods to both the full catalog and the Non-ML catalog. For the full catalog, we observe a strong correlation $L \sim Z^{2.8}$ after applying a flux limit of $4 \times 10^{-8}$ erg s$^{-1}$ cm$^{-2}$ keV$^{-1}$. After applying the method, we obtain $\tau = 3.84$, which indicates the presence of a strong luminosity evolution. We apply an identical procedure to the Non-ML catalog, this time obtaining a correlation $L \sim Z^{3.2}$, and $\tau = 3.35$. Both catalogs along with their truncation curves are shown in Figure \ref{fig:scatters}.

Following this, we proceed to determine the form of the evolution. Using Equation \ref{eq:Lumfit}, we calculate the local luminosity, given by 
$L_0 = L/g(Z)$, and use the Efron-Petrosian method to compute the associated value of $\tau$ by varying the value of the parameter $k$. For each sample, we compute the value of $k$ that gives $\tau=0$ and the associated $1\sigma$ range. For the full sample, we have $k=2.8\pm^{2.3}_{3.3}$.  For the Non-ML sample, we get $k=3.7\pm^{2.9}_{4.2}$. These results are shown in Figure \ref{fig:kvstau}.

\subsection{Luminosity Function}
After we compute the local luminosities by dividing by $g(Z)$, we  compute the local luminosity function. We model the cumulative luminosity function $\phi(L)$ using a smoothly broken power-law:

\begin{equation}
\phi(L) = \phi_0 \cdot \left(\frac{L}{L_0}\right)^{-\delta_1} \left[1 + \left(\frac{L}{L_0}\right)^{\delta_2 - \delta_1}\right]^{-1}
\end{equation}

The parameters for the fits are shown in Table \ref{tab:lumparams}. The figures for these fits are shown in the left panel of Figure \ref{fig:lumsforms}.

\vspace{0.5em}

\begin{table}[h!]
\centering
\caption{Luminosity Function Parameters}
\begin{tabular}{@{}lll@{}}
\toprule
\textbf{Parameter} & \textbf{Combined Catalog} & \textbf{Non-ML GRBs} \\
\midrule
$\phi_0$ & $3.08 \times 10^{2}$ & $1.72 \times 10^{2}$ \\
$L_0$ [$\mathrm{erg\,s}^{-1}$] & $3.56 \times 10^{50}$ & $1.56 \times 10^{50}$ \\
$\delta_1$ & $0.27$ & $0.31$ \\
$\delta_2$ & $1.35$ & $1.22$ \\
\bottomrule
\end{tabular}
\label{tab:lumparams}

\end{table}

\vspace{0.5em}

\subsection{Formation Rates}

\begin{table}[]
\centering
\caption{Best–fit parameters for Formation Rate}
\label{tab:doublebreak-fits}
\begin{tabular}{lcccccc}
\hline\hline
Catalog & $\a$ & $Z_{c1}$ & $\b_1$ & $Z_{c2}$ & $\b_2$ & $N_{0}$ \\
\hline
        Full   & 16.76 & 1.20 & 4.77 & 2.98 & 0.09 & 1.94 \\
    Non–ML & 11.70 & 1.39 & 3.57  & 2.81 & 0.09 & 2.68 \\
\hline
\end{tabular}
\end{table}

Using the same $C^{-}$ method, we compute the formation rates of both catalogs. For the fits, we adopt a \emph{double–break power law} as a phenomenological fitting function. We do not interpret each break as necessarily corresponding to a distinct physical transition; rather, the double-break form is used to flexibly describe the shape of the recovered cumulative formation-rate evolution. A simpler single-power-law form does not adequately capture the full curve, underestimating the comparatively shallow rise at low redshift and failing to reproduce the flattening toward the high-redshift end. We therefore use the double-break model as a descriptive parameterization of the reconstructed rate history:

\begin{equation}\label{eq:doublebreak}
    {\dot \sigma}(Z) \;=\; N_{0} \,
    \frac{Z^{\a}}{ \bigl(1 + (Z/Z_{c1})^{(\a-\b_1)}\bigr) \, \bigl(1 + (Z/Z_{c2})^{(\b_1-\b_2)}\bigr) },
\end{equation}
where $Z = 1+z$ is the redshift variable, $a$ controls the low–$Z$ rise,  
$Z_{c1}$ and $Z_{c2}$ are the two characteristic break redshifts with slopes $\b_1$ and $\b_2$,  
and $N_{0}$ is a normalization factor.

The best–fit parameters for both the Full and Non–ML catalogs are summarized in Table~\ref{tab:doublebreak-fits}. The fits themselves are shown in the right panel of Figure \ref{fig:lumsforms}.

\begin{figure}[]
    \centering
    \includegraphics[width=0.99\linewidth]{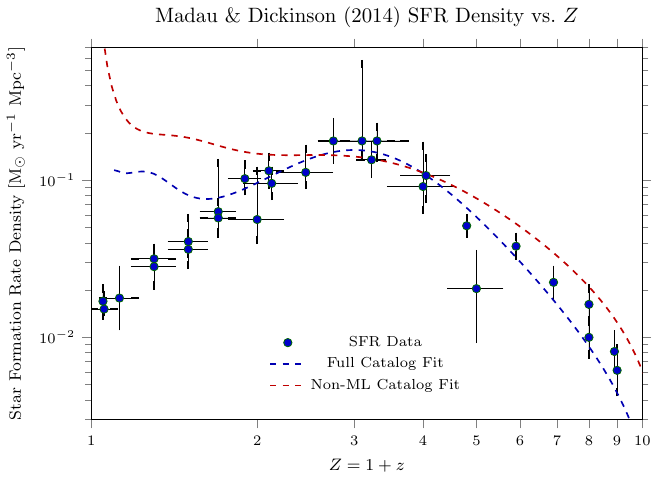}
    \caption{Comparison of the LGRB-derived formation rates with the star formation rate density from \cite{2014ARA&A..52..415M}. Green points with black error bars denote the observational SFR data, while the dashed blue and red curves correspond to the best-fit double-break models for the full and non-ML GRB catalogs, respectively.}
    \label{fig:sfr}
\end{figure}

After we determine the formation rate, we use the following equation to compute the density rate evolution, 

\beq
\label{densityrate}
{\dot \rho}(Z)=Z{d\sigma(Z)/dZ\over dV(Z)/dZ}.
\eeq
and compare it to the star formation rate in Figure \ref{fig:sfr}.

Because the ML-augmented catalog shows a reduced low-redshift excess relative to the non-ML catalog, it is important to consider whether this effect could be caused by a systematic bias in the ML redshifts. A direct EP reconstruction using only the overlap sample of GRBs with both spectroscopic and ML-predicted redshifts is not statistically meaningful, since this subset contains only about $20$ objects. In such a small sample, the associated sets used in the EP method are poorly populated, and the inferred luminosity evolution and formation-rate history become highly sensitive to individual bursts.

Nevertheless, the overlap sample provides a useful diagnostic of the direction of any possible redshift bias. In this subset, most of the ML-predicted redshifts are lower than the corresponding spectroscopic redshifts, rather than higher. Therefore, the reduced low-redshift excess in the ML-augmented catalog cannot be explained simply as the result of the ML estimator systematically shifting GRBs to higher redshift. Instead, the difference between the non-ML and ML-augmented rate histories is more plausibly connected to the redshift distribution of the added ML sample, which is concentrated primarily at intermediate redshifts, and to the small number of low-redshift GRBs added by the ML catalog.

\section{Robustness to $k$-Uncertainty}

To test whether our inferred luminosity function and formation-rate evolution depend sensitively on the precise value of the luminosity-evolution parameter $k$, we repeated the full $C^{-}$ reconstruction for the non-ML catalog using the best-fit value of $k$ and the two limiting values corresponding to the $1\sigma$ range, defined by $\tau=\pm 1$. For each value of $k$, we recomputed the de-evolved luminosities,
\begin{equation}
    L_0 = \frac{L(Z)}{g(Z)},
\end{equation}
and then recalculated the cumulative luminosity function, the cumulative formation-rate evolution, and the corresponding density rate evolution.

\begin{figure*}[t]
    \centering
    \includegraphics[width=0.48\linewidth]{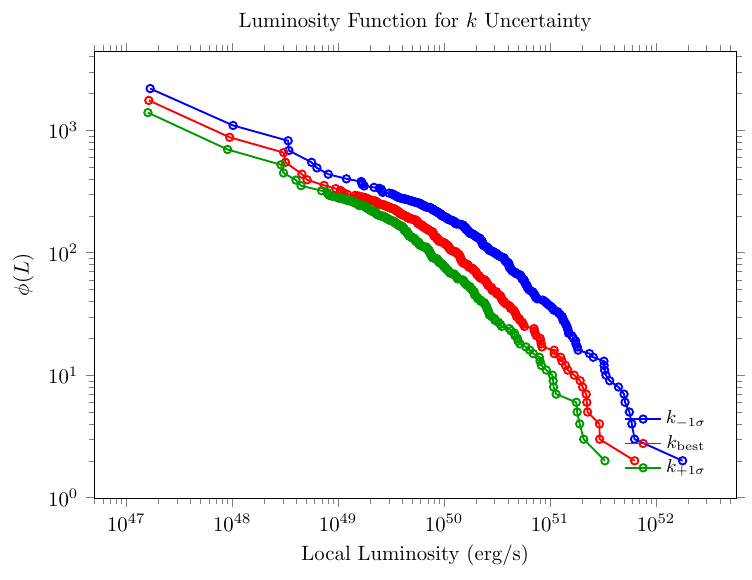}
    \includegraphics[width=0.48\linewidth]{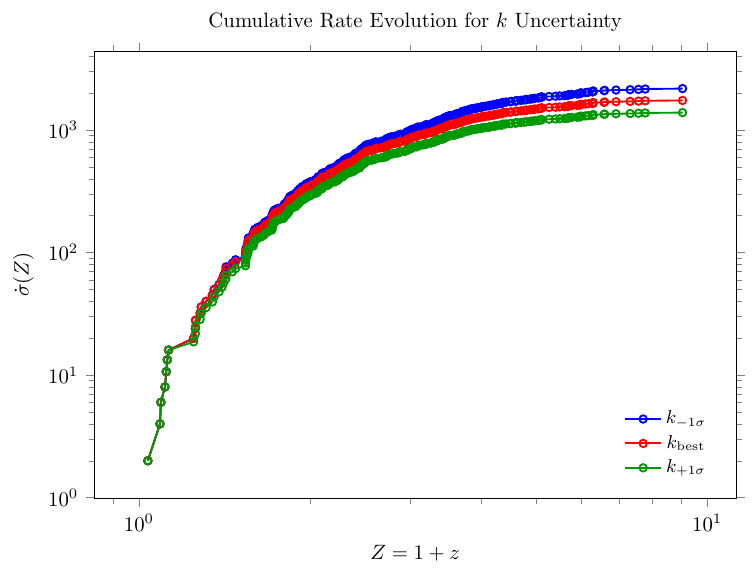}
    \caption{
    Robustness of the reconstructed luminosity function and cumulative formation-rate evolution for the non-ML GRB catalog to the uncertainty in the luminosity-evolution parameter $k$. 
    Curves are shown for the best-fit value of $k$ and for the two values corresponding to the $1\sigma$ EP range, defined by $\tau=\pm 1$. 
    The luminosity functions and cumulative formation rates retain nearly the same shape across the allowed range of $k$, differing primarily in normalization.
    }
    \label{fig:k-uncertainty}
\end{figure*}

The resulting luminosity functions and cumulative formation-rate curves are shown in Figure~\ref{fig:k-uncertainty}. Across the allowed $1\sigma$ range in $k$, both quantities retain nearly the same shape for the non-ML sample, with the main difference being an overall change in normalization. This indicates that the reconstructed luminosity function and cumulative formation-rate evolution are not strongly affected by the uncertainty in the luminosity-evolution correction.

\begin{figure}[t]
    \centering
    \includegraphics[width=0.99\linewidth]{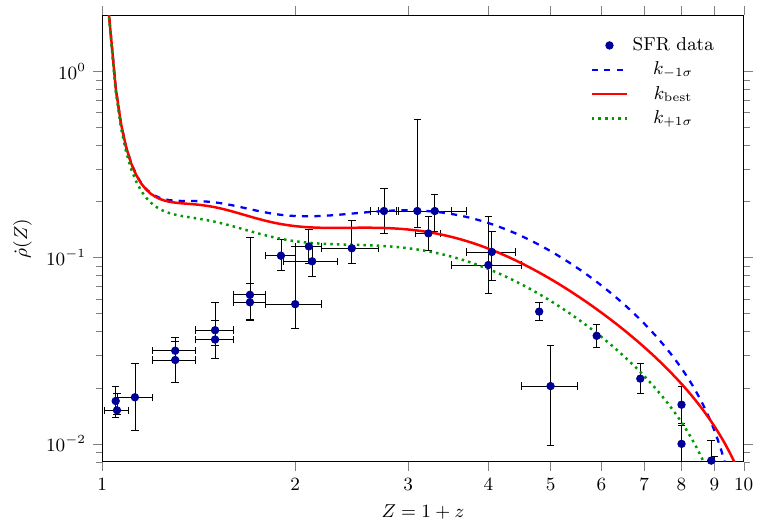}
    \caption{
    Density rate evolution inferred from the non-ML GRB catalog for the best-fit luminosity-evolution parameter $k$ and the two values corresponding to the $1\sigma$ EP range, overlaid on the observed cosmic star formation rate density. 
    Across the allowed range of $k$, the inferred density rate evolution shows the same low-redshift enhancement relative to the SFR. 
    At high redshift, the curves exhibit only small fluctuations and retain the same overall declining behavior.
    }
    \label{fig:k-uncertainty-sfr}
\end{figure}

We further compare the corresponding density rate evolutions with the cosmic SFR in Figure~\ref{fig:k-uncertainty-sfr}. The three curves display the same low-redshift behavior: in each case, the non-ML GRB density rate rises above the SFR at low redshift. At higher redshift, the curves show only small fluctuations relative to one another and follow the same overall declining trend. Thus, the observed low-redshift enhancement in the spectroscopic sample is not an artifact of adopting a single best-fit luminosity-evolution parameter, but instead remains stable across the $1\sigma$ range allowed by the EP analysis.

\section{Summary, Discussion and Conclusions}
\label{sec:concls}

In this study, our main goal was to investigate the luminosity and formation rate evolution of LGRBs, and the comparison of the latter with the cosmic SFR. While previous studies \citep{Petrosian_2015,2019MNRAS.488.5823L} have shared this objective, one of the improvements made in our work is the use of a more complete sample, including an increased sample of LGRBs with spectroscopic redshifts and accurate spectral models for a rest-frame luminosity comparison, and incorporating a considerable number of LGRBs whose redshifts have been determined via ML methods. 

In order to counteract the Malmquist-Eddington bias, common to all astronomical surveys, we use the nonparametric, nonbinning Efron-Petrosian methods to recast our catalog in terms of a local luminosity, which is corrected for luminosity evolution. From this, we determine  the (local) luminosity function and formation rate density evolution for both the entire LGRB  and the non-ML spectroscopic catalog. 

Our main conclusion is that the formation rate of LGRBs closely tracks the SFR beyond redshift for $z\geq 1.5$ for both catalogs. In the spectroscopic catalog, we see a gradually increasing deviation of the density rate from the SFR for $z<1.5$ reaching two order of magnitude deviation at the lowest redshift, confirming several earlier results summarized in \cite{Petrosian_2024}, and a more recent one by \cite{Khatiya_2025}. 
For the full catalog, the concordance with the SFR persists until $z \sim 1$, where the formation rate breaks and increases by a smaller factor than in the non-ML catalog. This reduced low-redshift excess is primarily associated with the redshift distribution of the added ML sample, which contributes relatively few low-redshift LGRBs.%

\footnote{One obtains similar discrepancies in LGRB rate when the samples are divided into low and high luminosity parts \citep{2011ApJ...739L..55B,2015ApJ...807..172N}, which, because of the observational selection bias, separates the sample into high and low $z$ samples.}

Although the overlap sample of GRBs with both spectroscopic and ML-predicted redshifts is too small for an independent EP reconstruction, most ML-predicted redshifts in this subset are lower than their spectroscopic counterparts. Thus, the reduced excess in the ML-augmented catalog is unlikely to be caused simply by a systematic upward shift in the ML redshifts.

In view of the recent discovery of the association of $\mathrm{GRB}211211\mathrm{A}$ and $\mathrm{GRB}230307\mathrm{A}$ with  kilonovae, \citep{Rastinejad2022Natur.612..223R,Mei2022Natur.612..236M,Levan2024},  \cite{Petrosian_2024} suggest that the progenitors of a significant fraction of the observed low $z$ excess of LGRBs could be NS-NS or NS-BH merger events, increasing the number of expected gravitational wave emitting sources.


However, several  factors can contribute to the  discrepancy between the SFR and LGRB rates at low redshift, such as bias from sample incompleteness, uncertainties arising from redshift determination, which requires accurate localization, and the  rapid optical/infrared afterglow follow-up, which can all affect the redshift distribution and rate evolution.


On the other hand,  \citet{2022MNRAS.513.1078D} identified a Gaussian-like excess in the LGRB formation rate at $z<1$, independent of sample completeness, which points toward a genuine physical origin of the low-$z$ excess.

At high-z, where the  metallicity is expected to be lower and stellar winds are less efficient, massive stars tend to retain their outer envelopes \citep{2006ARA&A..44..507W, 2005A&A...442..587V, 2022A&A...666A..14S}, preserving more angular momentum and increasing the likelihood of producing a collapsar and LGRB. 
In addition, a flattening of the stellar initial mass function  at high-$z$ may contribute to the enhanced LGRB rate  \citep{1998MNRAS.301..569L,2008ApJ...674...29V}. These could further enhance the significance of the low $z$ excess.

In future studies we hope to investigate the significance of these results in shedding light on the progenitors of LGRBs and their consequence for the rate of gravitational waves.

\section{Acknowledgements}
 M.G.D. acknowledges the support by JSPS Grant-in-Aid for Scientific Research (KAKENHI) (A), Grant Number
JP25H00675

\newpage

\bibliographystyle{aasjournal}
\bibliography{citations}
\end{document}